\documentclass[10pt]{article}

\usepackage[letterpaper]{geometry}
\usepackage{hicss51}
\usepackage{times}
\usepackage[none]{hyphenat}
\usepackage{url}
\usepackage{latexsym}
\usepackage{minted}
\usepackage{indentfirst}
\usepackage{graphicx}
\graphicspath{{images/}}

\usepackage{tikz, amsmath, pgf, paralist, xfrac, booktabs, multirow, breqn, tabularx}

\usepackage[normalem]{ulem}
\usetikzlibrary{positioning}

\title{Is Telehealth Better Used to Treat Patients or Help Other Physicians Treat Patients? An Agent-Based Modeling Study of Healthcare Provision}

\author{Michael Chary\\
  Weill Cornell Medical College \\
  {\underline{ mic9189@med.cornell.edu}} \\}

\date{}

\begin{document}
\maketitle
\begin{abstract}
Telehealth, the delivery of medical care remotely, is hoped to increase access to specialty services and improve health care utilization. Physicians can provide telehealth to each other (e.g. specialist to generalist) or to patients. Specialists often treat complex patients who can be adequately cared for only in academic hospitals, suggesting that providing specialty services via telehealth will reallocate rather than reduce system utilization. Here I use agent-based modeling to simulate telehealth’s effects on clinical outcomes and system utilization in medical toxicology. I found that toxicologist-physician consultation increased patient health and decreased cost. Toxicologist-patient telehealth increased cost and system utilization but did not improve health. The effects were sensitive to patient complexity and the clinical efficacy of the toxicologist. Within the limitations of using simulated data and an incomplete model, these results suggest that telehealth is more cost-effective when used to provide toxicologist access to general physicians than to the public.

\end{abstract}
\pagenumbering{gobble}
\section{Introduction}
 The remote provision of medical care, telehealth, is hoped to increase access to healthcare, for example by providing more specialty services such as neurology or cardiology to remote locations or outside traditional work hours. Telehealth can be provided between healthcare providers (physician-physician) or a healthcare provider and a patient (physician-patient). Earlier access to specialized care may prevent costly complications and preserve quality of life. However, specialists are likely to consider only causes within their expertise, raising the possibility that the effects of telehealth depend on whether the service is provided to a patient or a generalist physician. 

Medical toxicology is the medical specialty that diagnoses and treats the deleterious effects of substances on the body (\textit{e.g.} venom from animals, poisonous plants, exposure to caustic cleaning chemicals, or drug overdose). Toxicology is unique among medical specialties because it provides its services primarily via telehealth and diagnoses can often be made without specialized equipment, relying primarily on collateral information and physical exam findings, many of which can be elicited remotely. Medical toxicology, thus, provides a unique opportunity to ground a model of the relative efficacy of different points of healthcare access in real and relevant data.
  
Agent-based models (ABMs) have been used to model complex systems where controlled experimentation is not feasible.  An ABM of the delivery of toxicologic care provides a unique opportunity to explore the relative efficacy and cost-effectiveness of various points of access to care. We chose ABM over other modeling approaches because the components of an ABM readily correspond to parts of the real-life healthcare delivery system. 
 
\subsection{System for Delivering Toxicologic Care \label{ssec:history}}
The Industrial Revolution and urbanization led to increased exposure to cleaning agents (\textit{e.g.} bleach, ammonia) and workplace chemicals (\textit{e.g.} leaded gasoline\cite{zayed1994occupational}, kerosene\cite{morton1998hawaii}, asbestos\cite{selikoff1965relation}) in toxic amounts. In 1953, the first US poison control center (PCC) was established in Chicago, reflecting a growing recognition that poisonings (\textit{e.g.} drug overdoses, children getting into the medicine cabinet or cleaning agents under the sink, adverse reactions to new therapies) were occurring and that specialized knowledge was required to diagnose and treat them\cite{burda1997nation}. Initially, PCCs were funded by state governments and private companies and staffed by one toxicologist who oversaw pharmacists with additional training in toxicology. After recognizing Poison Control as a public service and rounds of consolidation, most PCCs today are funded by state governments and provide free consultation to physicians and the general public. 
  
To the public PCCs provide general information on the safe use of substances and advise when to seek further medical care, reducing the utilization of healthcare resources.  For example, most ingestions in young children are emotionally charged events, but rarely require medical intervention if the substances and amounts ingested can be verified.  To physicians, PCCs provide consultation for the definitive diagnosis and treatment of life-threatening poisonings, areas the general physician is unlikely to be familiar with. 

 Historically, a central question for the toxicologist was whether observation in a hospital setting (Emergency Department, Inpatient Ward, Intensive Care Unit) provides additional value. Because the vast majority of poisonings do not require acute treatment, toxicologists have demonstrated most value in minimizing the testing and interventions done on mildly poisoned patients. Patients in whom the substance and amount are reasonably certainly known and who have no conditions that deplete physiologic reserve or alter metabolism may often be safely observed at home in the company of a friend or family member.  Depending on the nature or amount of substance ingested, some patients may benefit from treatment in the Emergency Department without admission to the hospital. Any person who presents to an Emergency Department is screened and evaluated for all likely emergent conditions, often leading to additional uninformative and costly tests. Most tests to determine blood concentrations of toxins take days to weeks to provide results. In severe poisonings, prompt treatment by a toxicologist may prevent avoid clinical deterioration, restoring the patient's health enough to require admission, but not admission to the Intensive Care Unit.  

As more people live longer and with multiple chronic illnesses, toxicologist-patient consultation may be less able to decrease health care utilization. Complex patients are more likely to need a complete medical evaluation, which cannot be done over telehealth. Health care centers (hospitals, nursing homes, long-term rehabilitation and psychiatric facilities) make up an increasing portion of the calls to PCCs\cite{lai20062005}. The general public is increasingly turning to the Internet. Decreasing funding for PCCs further motivates the search for alternative ways to provide toxicological care. In tandem, more hospitals have toxicologists on staff than, before, increasing the feasibility of toxicologist-physician in-person consultation.

\subsection{Approaches to the Economic Analysis of Healthcare Systems}\label{sec:econ}
 This section discusses \emph{cost-effectiveness} and \emph{incremental cost-effectiveness} as they relate to the evaluation of health care systems.
 
 Cost-effectiveness refers to the ratio of the cost of providing a service to its outcome. Outcome is usually quantified as the patient's improvement in quality of life or life-expectancy. It is typically expressed as QALYs (quality-adjusted life years), a measure that combines the gain in longevity and quality of life into a single number\cite{neumann2016cost, loomes1989use}.  The incremental cost-effectiveness ratio (ICER) calculates the ratio of the differences in costs to differences in outcomes for two interventions.
 
The national health agency in the United Kingdom (NICE) and pharmaceutical agencies in America use QALYs (Qualify-Adjusted Life Year) and ICER when evaluating different therapeutics and interventions\cite{ogden2017qalys, neumann2009united, neumann2018qalys}. The median ICER for interventions across three insurance companies was $\$12,500$ to $\$32,200$ per QALY in 2003 US dollars\cite{king2005willingness}. In this study payments were reported as bundles, not itemized. A rule-of-thumb is that an intervention is cost-effective if it is less than $\$50,000$ pre QALY, although the Center for Medicaid and Medicare does not use any explicit threshold when setting reimbursement rates\cite{weinstein2008much}. The median cost-effectiveness of ICU stays is reported to be $\$4,100$ per QALY, varying from $\$640$ for alcohol intoxication to $\$ 30,625$ for kidney failure\cite{sznajder2001cost}. Implementing a team to optimize antibiotic use (antibiotic stewardship team), decreased the frequency and duration of systemic infections in the Intensive Care Unit (ICU), with an ICER of $\$2,637$ per QALY\cite{scheetz2009cost}. However, ICERs may not fully capture all changes in costs and depend on accurately assessing improvement in quality of life.

\subsection{Current State of Telehealth} 
Telehealth began with providing neurology expertise in caring for strokes to remote areas. It now also helps patients in remote locations manage chronic conditions without needing to always come to a hospital or physician's office. A substantial barrier in some specialties to using telehealth is the difficulty in building trust (\textit{e.g.} psychiatry) or performing a physical examination (\textit{e.g.} surgery). However, psychiatrist-physician telepsychiatry has proven useful in assessing imminently life-threatening conditions, such as suicidality, homicidality, or acute psychosis. Wearable devices partially offset the limitations of telehealth physical examination\cite{skolnik2016teletoxicology}. Insurance companies rarely reimburse for telehealth, thus favoring uses of telehealth that reduce the costs associated with delivering services that are reimbursed.

The ICER associated with telehealth for managing depression in patients with established depression was \textsterling 132, 630\cite{Dixon2016} and \textsterling 92,000 for managing chronic medical conditions (high blood pressure, diabetes, high cholesterol) in patients who also had access to office-based care\cite{henderson2013cost}. A telehealth short-term intervention in addition to primary care but in lieu of cardiology referral had an ICER of \textsterling 10, 392\cite{dixon2016cost}.

\subsection{Prior Work: Cost-Effectiveness of Toxicological Care \label{ssec:prior-work}} 
Poison Control Centers, historically, improve health by reducing unnecessary treatments and tests, which, in turn, reduces health care system utilization and costs\cite{woolf2011preserving}. A cost-effectiveness analysis of PCCs reported that toxicologist-physician consultation via PCCs decreased cost, no matter the poisoning (ICER $\$50,000$ to $\$250,000$)\cite{harrison1996cost}. Another analysis of PCCs calculated that for approximately $\$825 $ per case, PCCs reduced visits to the Emergency Department by 350,000 visits per year and the number of hospitalizations by 40,000 per year\cite{miller1997costs}. An analysis of rural areas calculated that every 43 calls to a PCC saved 1 hospital admission to the hospital, worth $\$8,000$\cite{zaloshnja2006potential}.  

In-person toxicology consultation reduces duration of admissions and increases hospital revenue, but its efficacy varies with the complexity of the poisoning. A retrospective study of $1,534$ mildly poisoned and $2,227$ severely poisoned patients admitted to one hospital system in Arizona found that in-person toxicology consultation decreased length of hospitalization by $0.5$ days for mildly poisoned patients and $1$ day for severely poisoned patients; with an increase in $33$ lives saved per $1,00$ and  roughly $\$1,110$ fewer healthcare costs per patient\cite{curry2015effect}. Another retrospective study of $123,000$ patients with exposure but not necessarily clinical manifestations of poisoning found that patients admitted to a dedicated toxicology service had $0.87$ days shorter length of hospitalization and the hospital received an average of $\$1,800$ more per patient encounter\cite{king2019effect}. A comparison of $88$ patients before and $88$ patients after starting an in-person toxicology consultation service had no effect on clinical outcomes\cite{Clark1998}. 

\subsection{Prior Work: Agent-Based Modeling}
Agent-based models (ABMs) are kinds of finite dynamical systems that simulate the actions and interactions of autonomous agents over time. They have been used to simulate the impact of adding lanes to a highway on traffic flow, the interdependence between predator and prey populations, and the impact of proposed zoning laws on urban traffic flow and pollution\cite{gilbert2019agent}. More recently, they have been used to model the dependence of the dynamics of COVID-19 infection to the rate of vaccination or social distancing\cite{hoertel2020stochastic}. ABMs are typically composed of one or more types of agents, rules specifying how each agent evolves and interacts with other agents, and, optionally, spatial constraints, \textit{e.g.} when modeling traffic flow or spatially heterogeneous quantities, like wealth or rainfall. When the agents interact on a two-dimensional grid, the simulations resemble cellular automata.

In analyzing healthcare systems, ABMs have been used to identify the most cost-effective screening interval for eye complications of diabetes\cite{day2014sensitivity}, the influence of geography on recommendations to reduce risk for cardiovascular disease\cite{li2015advancing}, and the influence of peer networks in modulating the effectiveness of interventions to decrease adolescent obesity\cite{zhang2015leveraging}. 

\subsection{Goals of This Study \label{ssec:goals}} The goal of this study was to determine the cost-effectiveness of each point of access to toxicological care (toxicologist-physician, toxicologist-patient via PCC, toxicologist-patient via telehealth).  A secondary goal was to determine the sensitivity of cost-effectiveness to clinical complexity and the efficacy of clinical care. It was assumed that the system for delivering toxicological care aims to maximize health improvement and minimize resource use.

\section{Methods \label{sec:methods} }
 We created an agent-based model using the \textit{mesa} package in Python\cite{masad2015mesa}. Supporting code, figures, and documentation are available at the GitHub Repository, \textsc{abm-pcc}. The model consists of three mutually exclusive types of agents: Patients, Physicians, and Toxicologists (Table \ref{tab:agents-in-mode}). In this model the term \emph{Patient} includes people at home who self-treat, use toxicologist-patient telehealth, or call PCC as well as those presenting to a hospital.
 Patients become poisoned and, depending on their willingness, seek care in a hospital or virtually. Patients who seek care in a hospital are treated by a Physician and, if the presentation is severe enough, a Toxicologist whom the Physician consults. Patients who do not seek care in a hospital try to treat themselves at home by looking for information on the Internet, calling Poison Control, or directly contacting a Toxicologist via telehealth. 
 
 The central differences between toxicologist-patient consultation via PCC and via telehealth is that via telehealth patients interact with the same toxicologist repeatedly, but PCC consultation is free. Telehealth consultation costs more but improves health more than PCC consultation. The greater effectiveness of telehealth reflects the assumption I made that repeated interactions or the use of video or wearables would allow toxicologists to perform a fuller evaluation via telehealth than PCC. Looking information up on the Internet is free but sometimes provides harmful information. This models the effect of inaccurate or misleading information online as well as the difficulties inherent in self-diagnosis. Contacting PCC will not provide harmful information. But, if the patient cannot fully describe the poisoning, the patient may be inappropriately referred to a hospital or instructed to stay home. Inappropriate referral signifies advising a patient to go to a hospital even if the patient is unlikely to benefit from medical care, thereby consuming health care resources with no improvement in health. Inappropriate keeping at home refers to advising patients whose illness requires treatment that can only be accomplished in the hospital to stay at home. Instructing the person to stay at home risks delaying care and, possibly, worse health outcomes. 
 
 The propensities to access each point of care or become poisoned are modeled as thresholds over uniform variables. For each propensity all Patients have the same threshold. At each step the Patient realizes the propensity (\textit{e.g.} calls PCC) if a random number is below that threshold. All random numbers were chosen by the built-in random number generator provided in Python 3.9.1 over $\left[0,1\right]$. 
 
 Table \ref{tab:sources-for-parameters} details the parameters and, when available, empirical work used to choose specific values. In Table \ref{tab:sources-for-parameters}, the term value refers to the numerical value used in the simulation. All fractions are between $0$ and $1$, health effect was between $-100$ and $100$. The relative treatment effect, a ratio, had a lower bound of $1$. These values were used in conjunction with the agent behavior shown in in Table \ref{tab:agents-in-mode}.
 
\begin{table}[]
\caption{Literature sources for parameters.}
\label{tab:sources-for-parameters}
\begin{tabularx}{\columnwidth}{@{}Xll@{}}
\toprule
Variable & Value & References \\ \midrule
Fraction of population with poisoning                           & 0.1  &  \cite{mowry20152014, gummin20202019}\\
With fatal poisoning                     & 0.005 &  \cite{mowry20152014, gummin20202019, giurca2020time, kordrostami2017forensic} \\
With nonfatal poisoning, low clinical complexity & 0.8  &  \cite{gummin20202019, townsend2001substances, kordrostami2017forensic} \\
With nonfatal poisoning high clinical complexity          & 0.2            &      \cite{gummin20202019, townsend2001substances}      \\
Relative treatment efficacy of toxicologist vs generalist                & 2.5  & \cite{warrick201829} \\
Health effect of misinformation online & -20            &   \cite{brush2004monoamine,chary2020geospatial, karami2018trends}         \\ \bottomrule
\end{tabularx}
\end{table}

\subsection{Experimental Design}
Each simulation contained $1,000$ Patients, $100$ Physicians, and $10$ Toxicologists. The results reported in the \textbf{Results} section show the median result. 

Table \ref{tab:experimental-design} describes the scenarios investigated. A plus sign indicates the availability of that line of service and minus sign, the lack. We chose this factorial design to explore the effects of different lines of service and their interactions. Each scenario was run in replicate $10$ times. To account for the sensitivity of results to clinical complexity, each replicate was run $10$ times, varying clinical complexity with each run from $0$ to $1$ in steps of $0.1$, leading to $10\cdot 10=100$ simulation runs.

\begin{table}[b]
\centering
\caption{Simulation plans. Each column denotes a simulation plan.(+) denotes presence of factor, (-) absence. TELE, toxicologist-patient telehealth, TOX, toxicologist-physician telehealth; PCC, toxicologist-patient telehealth with no follow-up, \textit{i.e.} Poison Control Center}
\label{tab:experimental-design}
\begin{tabular}{@{}lllllllll@{}}
\toprule
TELE & + & + & + & + & - & - & - & - \\ 
TOX  & + & + & - & - & + & + & - & - \\
PCC  & + & - & + & - & + & - & + & - \\ \bottomrule
\end{tabular}
\end{table}

We simulated clinical complexity as a single parameter that varied from $0$ to $1$. Clinical complexity refers to factors extrinsic to the poisoning. For example, a snakebite could inject venom that causes extreme swelling of the limbs and a predisposition to bleed. These effects could be more dangerous in someone being treated with blood thinners for another disease or with a hereditary disposition to bleed easily. It also refers to the effect of social determinants of health, logistical issues (\textit{e.g.} delays in seeking care), or differences in amounts of substance ingested.

\begin{table*}[t]
\caption{Behavior of Agents in Model. Brackets indicate ranges for variables.}
\label{tab:agents-in-mode}
\begin{tabularx}{\textwidth}{@{}lXXX@{}}
\toprule
 &
  Properties &
  Actions &
  Rules \\ \midrule
Patient &
  \begin{tabular}[c]{@{}l@{}}Health {[}0,100{]}\\ Complexity of Care {[}0,1{]} \\ Bill {[}0,$\infty${]}\\ Severity of Poisoning {[}0,100{]}\\ Willingness to Seek Care {[}0,1{]} \end{tabular} &
  \begin{tabular}[c]{@{}l@{}}Call PCC\\ Present to Hospital \\ for Treatment\\ Misinterpret Severity of Poisoning\end{tabular} &
  \begin{tabular}[c]{@{}l@{}}Die if health is 0\\ Seek care no matter what\\ if health is less than 50\\ Be billed if seek care\end{tabular} \\ \midrule
Physician &
  \begin{tabular}[c]{@{}l@{}}Treatment Effect\end{tabular} &
  \begin{tabular}[c]{@{}l@{}}Treat Patient\\ Consult Toxicologist\\ Bill\end{tabular} &
  \begin{tabular}[c]{@{}l@{}}Increase Patient's Health by 10\\ Bill Patient for 100\\ Consult Toxicologist for \\ High Complexity Cases\end{tabular} \\ \midrule
Toxicologist &
  \begin{tabular}[c]{@{}l@{}}Treatment Effect\end{tabular} &
  \begin{tabular}[c]{@{}l@{}}Treat Patient by Responding \\ to Physician Consult\\ Bill\end{tabular} &
  \begin{tabular}[c]{@{}l@{}}Multiply Physician's Effect by 2.5\\ Multiply Physicians Bill by 2.5\\ Provide Information Via PCC\end{tabular} \\ \bottomrule
\end{tabularx}
\end{table*}

\begin{figure}
    \centering
    \begin{tikzpicture}
    \node (person) {Patient};
    \node[below =of person] (poisoned) {Poisoned};
    \node[below = 1em of poisoned] (juncture){.};
    \node[left= of juncture] (die) {Die};
    \node[right = of juncture] (home) {Not seek care};
    \node[below left=of juncture] (internet){Internet};
    \node[below=of juncture](hospital){Hospital};
    \node[below right = of juncture] (pcc) {PCC};
    \node[right = 0.5cm of pcc](telehealth){Telehealth};
    
    \node[below = of hospital](physician) {Physician};
    \node[below = of physician](toxicologist) {Toxicologist};
    
    \draw[->, thick] (juncture) -- (home);    
    \draw[->, thick] (person) -- node[left]{$10\%$} (poisoned);
    \draw[thick] (poisoned) -- (juncture);
    \draw[->, thick] (juncture) -- node[above]{$5\%$} (die);
    
    \foreach \locale in {internet, hospital, pcc}
    {\draw[->,thick] (juncture) -- (\locale);}

    \draw[<->, thick] (hospital) -- (physician);
    \draw[<->, thick] (pcc) -- (hospital);
    \draw[->, thick] (physician) -- (toxicologist);
    \draw[->, thick] (pcc) -- (toxicologist);
    
    \draw[<->,thick] (telehealth) -- (toxicologist);
    \draw[<->,thick] (juncture) -- (telehealth);
    
    \end{tikzpicture}
    \caption{\textbf{Interaction Diagram of Agents.} PCC, Poison Control Center. Arrows indicate possible paths. Bidirectional arrows indicate repeated interactions.}
    \label{fig:abm}
\end{figure}
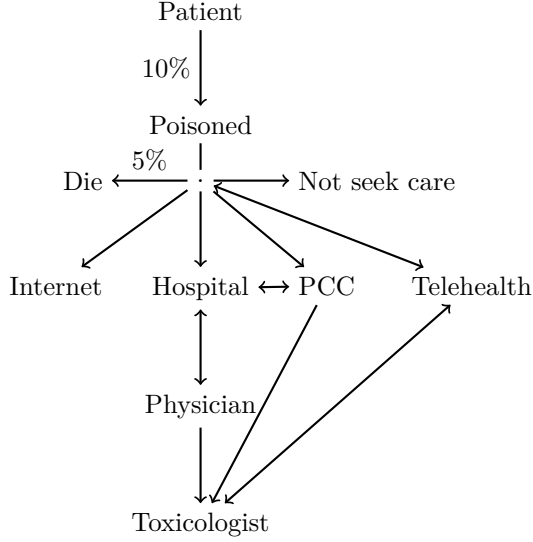

\section{Results \label{sec:Results}}

\begin{figure}
    \centering
    \includegraphics{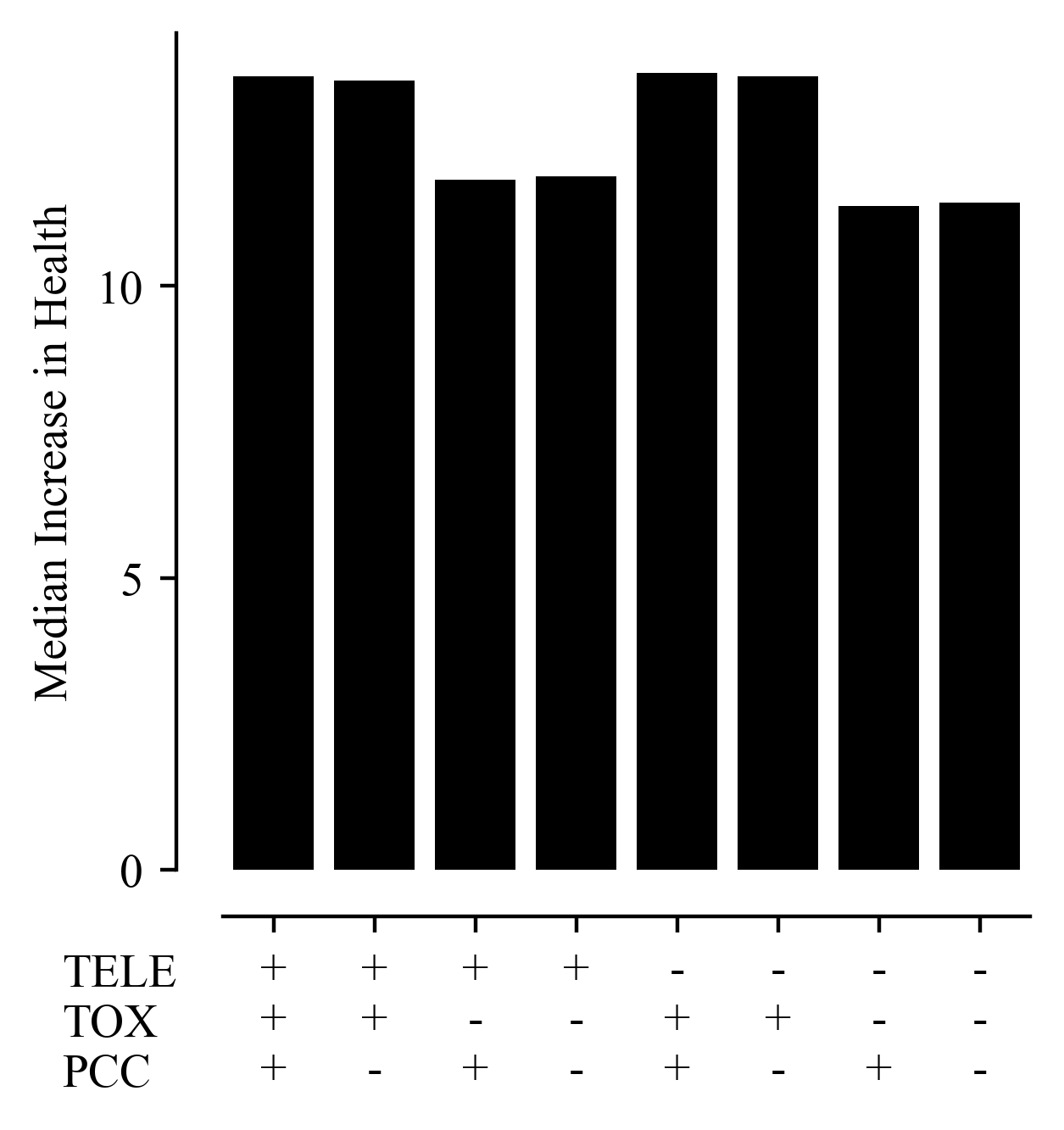}
    \caption{\textbf{Effect of Point of Access to Toxicological Care on Health Improvement.} Y-axis shows improvement in health. X-axis shows availability (plus) or lack (minus) of point of access. \textsc{TELE}, physician-patient telehealth; \textsc{TOX}, physician-physician telehealth; \textsc{PCC}, physician-patient telehealth with no follow-up (i.e. poison control center). Total increase calculated as the area under the curve, \textit{e.g.} in the panels in Figures \ref{fig:health-cost-complexity} or \ref{fig:telehealth}.}
    \label{fig:summary-health}
\end{figure}

\begin{figure}
    \centering
    \includegraphics{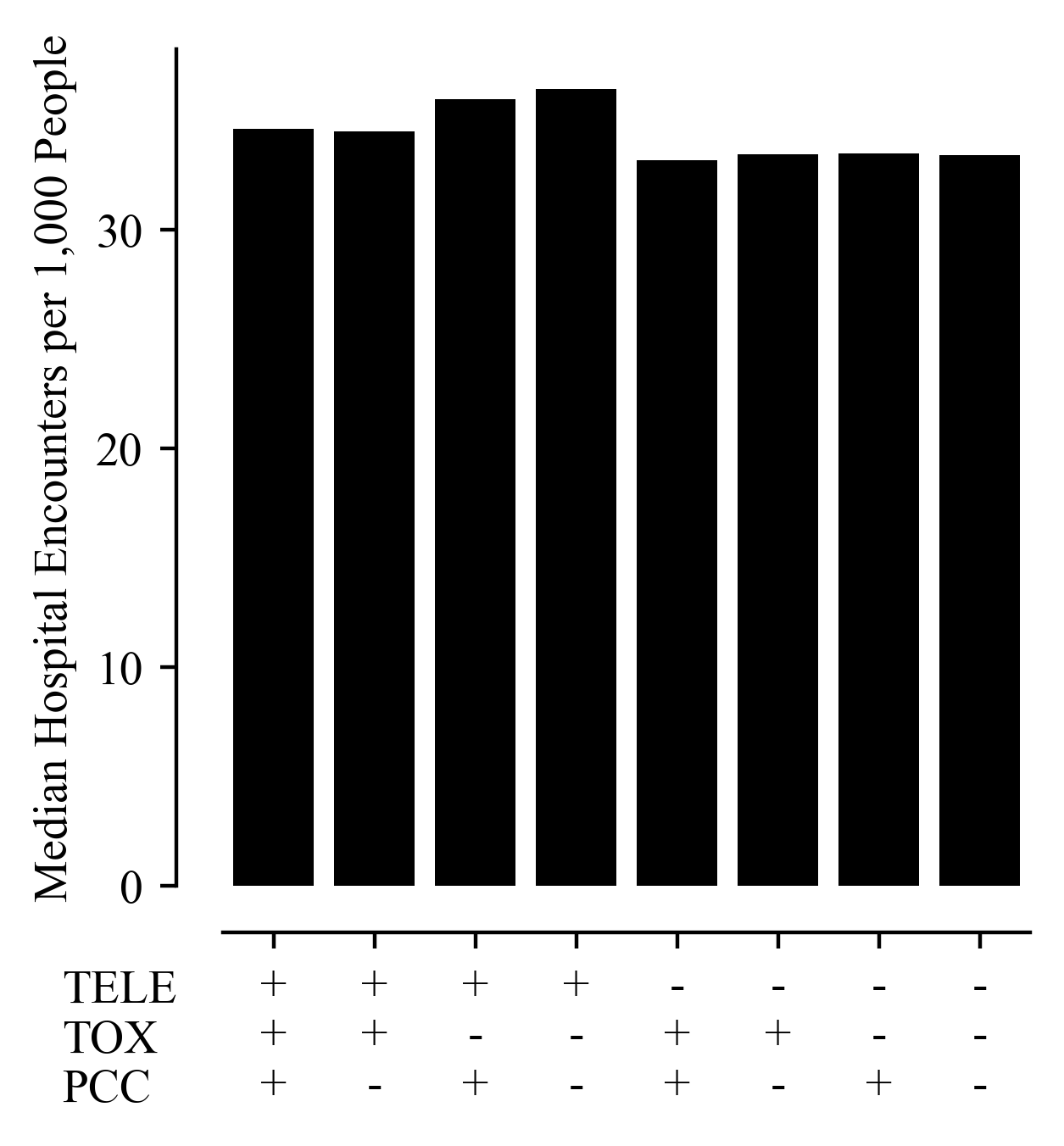}
    \caption{\textbf{Effect of Point of Access to Toxicological Care on Total Number of Patients Admitted.} Axes same as for Figure \ref{fig:summary-health}.}
    \label{fig:summary-admitted}
\end{figure}

\begin{figure}[t]
    \centering
    \includegraphics{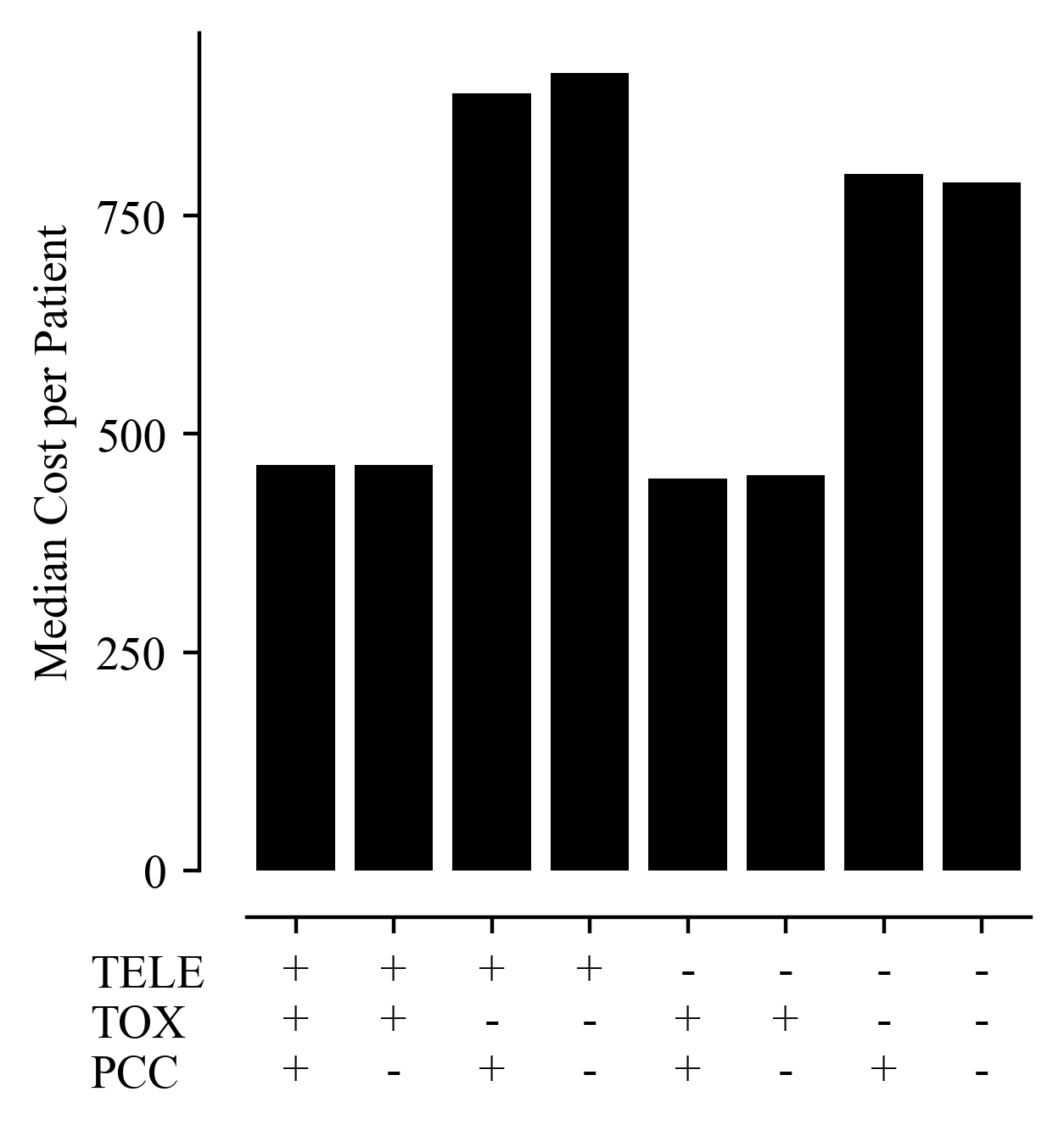}
    \caption{\textbf{Effect of Point of Access to Toxicological Care on Cost.} Axes and calculation same as for Figure \ref{fig:summary-health}, except that y-axis is logarithmic here.}
    \label{fig:summary-cost}
\end{figure}

Figures \ref{fig:summary-health} to \ref{fig:summary-cost} illustrate the relationships between improvement in health (Figure \ref{fig:summary-health}), hospital admissions (Figure \ref{fig:summary-admitted}), cost (Figure \ref{fig:summary-cost}) and which points of access for toxicological service were available. 

Figure \ref{fig:summary-health} describes the median increase in health per patient. Health was quantified as a number between $0$ (death) and $100$ (full health), with $50$ denoting a critical illness. We considered a mild poisoning to subtract $20$ from a patient's health. A severe one subtracted $40$. Health increased when no toxicological care was available because generalist care was still available and self-treatment was possible. PCC care provided no benefit beyond toxicologist-patient consultation via telehealth. Access to toxicologist-physician consultation was associated with the biggest increase in health per patient, even though only $1\%$ of the population had poisonings severe enough to use this point of access.

Figure \ref{fig:summary-admitted} describes the median hospital encounters per $1,000$ Patients. These encounters decreased when the only point of access for toxicology care was toxicologist-physician consultation. This decrease reflects the lack of toxicologists referring patients in. Toxicologist-patient telehealth 
was responsible for more referrals to hospital than PCC was. This reflects more accurate recognition of complex or severe poisonings with toxicologist-telehealth than PCC. Care via toxicologist-patient telehealth partially offset the increase in hospital admissions when toxicologist-physician consultation was not available (Figure \ref{fig:summary-admitted}). 

Figure \ref{fig:summary-cost} describes the median cost per patient. Toxicologist-physician consultation allowed an approximately $20\%$ reduction in cost Toxicologist-patient telehealth increased costs, reflecting how some Patients used telehealth, a paid service, over PCC. Toxicologist-patient PCC reduced costs, although by less than telehealth increased them.  The overall cost decreased despite the additional fees of providing toxicologist-physician consultation because the consultation reduced length of stay by $2-4$ timesteps (equated in this analysis with days). At each timestep, the physician billed $100$ to increase health by $10$ and the toxicologist billed $150$ to increase health by $25$, leading to an ICER of $\sfrac{150}{25}\Rightarrow 6$. The availability of PCC did not change cost, as expected, because this is a free service. 
 
\begin{figure}[h!]
    \includegraphics[scale=0.9]{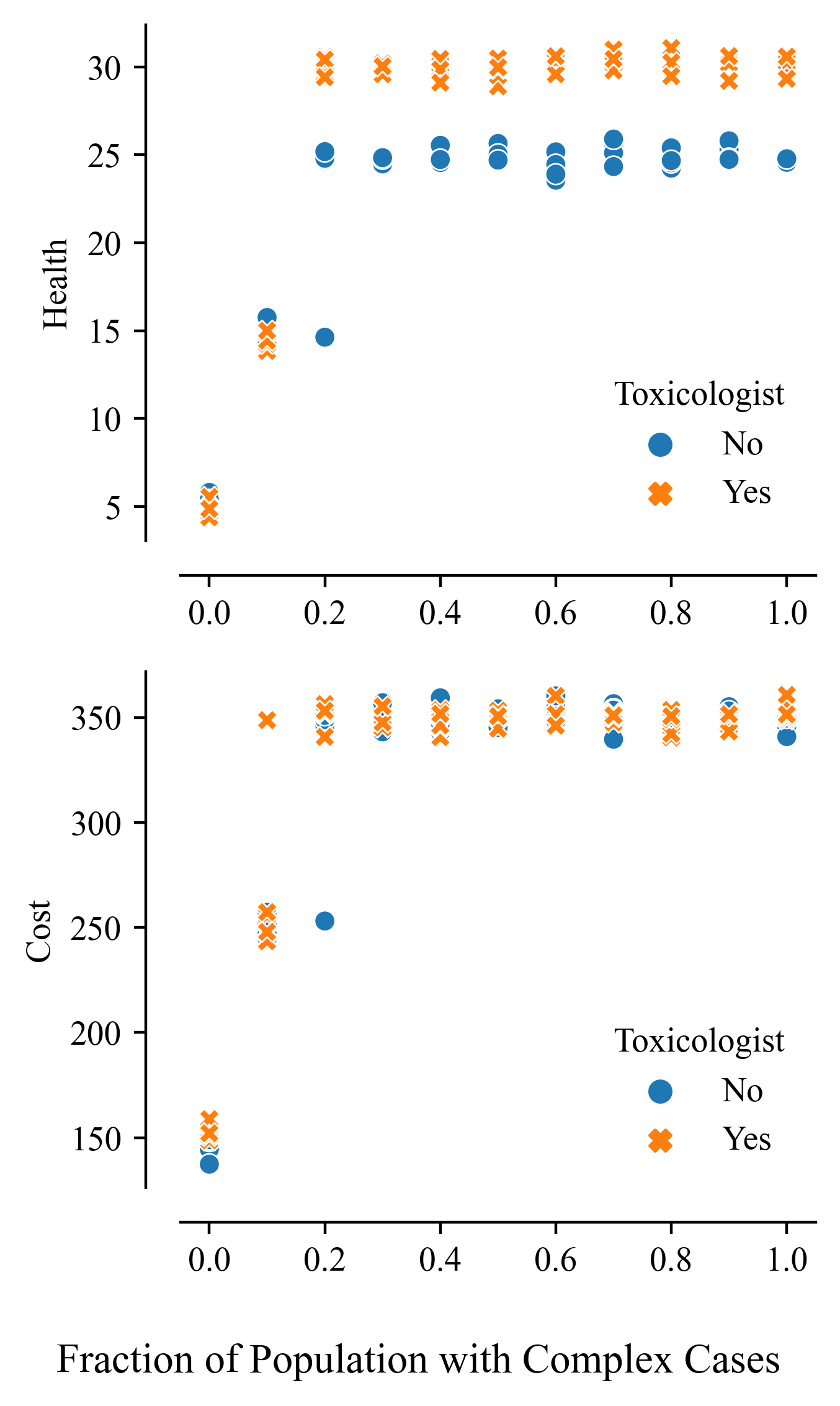}
    \caption{\textbf{Impact of Complexity and Toxicologist Care on Median Health (top) and Cost (bottom) of Treated Patients}. X-axes show complexity. Y-axes, health (top) or cost (bottom). Crosses represent simulations with toxicologist care available, circles, no toxicological care available. Each cross or dot represents one simulation.}
    \label{fig:health-cost-complexity}
\end{figure}

This model assumed that a fraction of the population would have clinically complex cases and present directly to the hospital. Figure \ref{fig:health-cost-complexity} displays the sensitivity of changes in health and cost per patient to clinical complexity. In general, a clinically complex patient is the more likely to require long hospitalizations and specialist consultation than a clinically simple patient. Toxicologist-physician consultation improved health with no systematic impact on cost (blue circles compared with orange dots).
The presence of any point of access to toxicological care did not change the shape of sensitivity function. Both cost and health increased and then saturated. The saturation of health reflects that patients were discharged from the hospital when their health was $\sfrac{90}{100}$ and that severe poisonings subtracted $40$ from a Patient's health. The saturation of cost reflects the trade-off between shorter length of stay, which reduced general physician billing, and the increased billing of the toxicologist-physician consultation. 

\begin{figure}
     \includegraphics{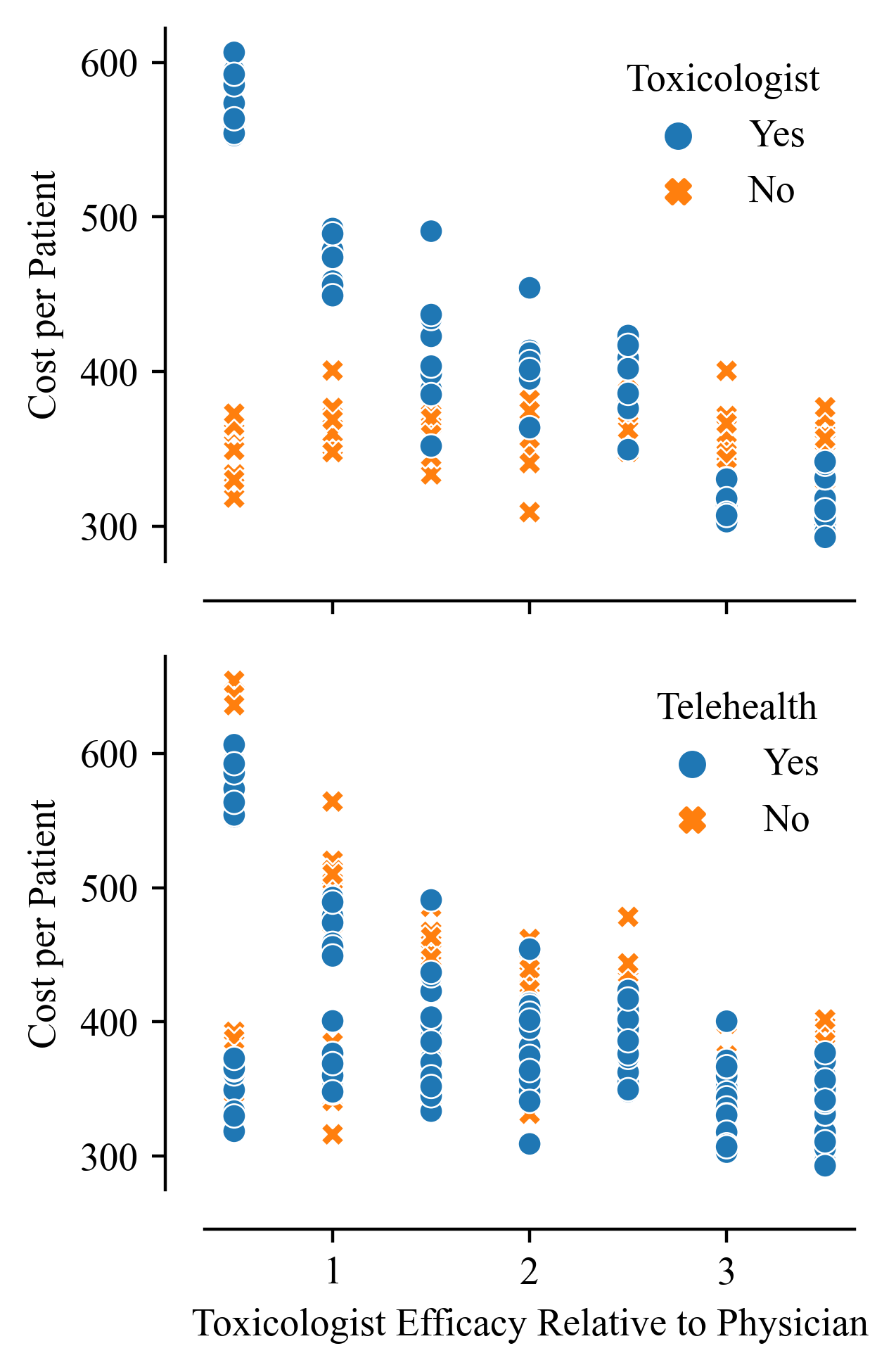}
     \caption{\textbf{Relationship between Clinical Efficacy and Overall Cost Per Patient for Toxicologist-Physician (top) and Toxicologist-Patient Telehealth (bottom) lines of service.} Legend indicates presence or absence of service}
     \label{fig:telehealth}
 \end{figure}

 This model assumed that toxicologists help general physicians treat complex poisoned patients more effectively than general physicians could on their own. Figure \ref{fig:telehealth} displays the sensitivity of cost to toxicologist efficacy. The x-axis displays the relative efficacy of the toxicologist to the general physician. A relative efficacy of $0.5$ means that for every unit health the general physician increases, the toxicologist increases an additional $0.5$ units, \textit{i.e.} has less half the relative effectiveness of the generalist. For both toxicologist-physician (top panel) and toxicologist-patient telehealth, cost per patient is a decreasing function of the toxicologist's clinical efficacy. The highest cost per patient occurs when telehealth is not available (but PCC and toxicologist-physician consultation are) and the toxicologist is half as effective as the general physician, prolonging length of stay in the hospital. Figures \ref{fig:summary-health} to \ref{fig:summary-cost} used a relative efficacy of $2.5$.

\section{Conclusions\label{sec:Conclusions}}
 The goal of this paper was to understand whether toxicologists improve health outcomes more cost-effectively when providing patient- or physician-facing points of access. A secondary goal was to determine the sensitivity of this cost-effectiveness to clinical complexity and efficacy. I created an agent-based model to capture the effects of toxicologist-physician consultation via PCC or telehealth, and toxicologist-physician in-person consultation. As demonstration of its validity, the model captures the previously described cost-savings of toxicological care and PCCs.

 The key finding of these simulations is that (the model's representation of) toxicologist-physician interactions increase health and decrease costs, while toxicologist-patient telehealth had the opposite effect. Toxicologists and general physicians working together improved health more than either working alone could. Recognizing that patients with more complex poisonings are more likely to seek care in hospital, the main implication of this work is that the primary way toxicologists improve health is by improving the care of complex patients. Toxicology-physician consultation had no effect on system utilization because the physician consulted the toxicologist after the patient was being treated in the hospital.
 
 The results presented here agree with previously reported empiric findings in toxicology and telehealth. A study of poison centers in Illinois divided patients into five levels of clinical complexity and found PCC consultation was associated with nearly $\$800$ increased costs in the least clinically complex patients but a savings of about $\$4,500$ in the most clinically complex patients\cite{friedman2014association}. Telepsychiatry for patients with established psychiatric diagnoses was associated with increased cost and hospitalization as compared to in-person care\cite{modai2006cost}, and the cost-effectiveness was dependent on practitioner experience with telehealth\cite{naslund2020economic}. 
  
  A criticism of ABMs is that their results depend on many parameters whose values are not empirically known and that specific results may strongly depend on initial conditions \cite{sorensen1976models} These concerns apply to most investigations of dynamical systems, especially around bifurcation points. Our approach is mindful of this pitfall and investigates only general dynamics and makes no conclusions about specific numerical results. We assessed the sensitivity of our results to clinical complexity and the magnitude of improvement in health that toxicologists bring. Another criticism is that ABMs do not give reproducible results\cite{dawid2019macroeconomics}. The results we present are the average of $100$ runs.
 
 This study is a starting point that demonstrates the utility of agent-based modeling in understanding the role of various points of access in healthcare systems. However, there are limitations to this study that future work can address. The model considered each patient to have only one medical problem and assumed that (1) there were no medical errors or side effects from treatment in the hospital, (2) in-hospital diagnostic accuracy was perfect, (3) treatment universally effective, (4) all improvements in health were equal, and (5) that no toxicologist-physician communication happened via telehealth. Our analysis of cost-effectiveness is sensitive to the relative magnitude of treatment effects of physicians and clinical complexity, although in an expected and plausible manner. I could find no empirical data for the absolute or relative values of clinical complexity. The model focuses on acute poisonings. It does not consider chronic exposures, as might happen in the workplace, in a specific location because of industrial pollution, or from failing to adjust medication doses as a patient ages or experiences organ failure. Nor does the model explicitly consider age, although it is implicit in the clinical complexity variable. This model does not differentiate between treatment in the ED, admission to a general hospital ward, or admission to the Intensive Care Unit. Toxicologist-patient telehealth may be cost-effective if it, similar to toxicologist-physician consultation, decreases the need the for Intensive Care Unit admissions even if it increases the overall need for admissions, \textit{i.e.} capturing people earlier in their illnesses. Admission to the Intensive Care Unit is approximately $10$ times more expensive than admission to the general ward, owing to the cost of increased monitoring, individual nursing care, and frequent use of life-saving procedures and equipment, such as mechanical ventilation. Indeed, ICU is more cost-effective for the most critically (yet not fatally) ill patients \cite{edbrooke2011implications}, suggesting that preventing admission to the ICU for cases that can be adequately treated on the floor with a toxicological consultation would be cost-saving. Using telehealth to provide physician coverage to ICUs in remote locations (often only staffed by nurses and physician assistants) was associated with an ICER of $\$25,392$ (2014 USD), driven by preventing transferring the $40\%$ most critically ill patients between ICUs.

 Limitations not withstanding, these results provisionally suggest that for acute poisonings the largest and most cost-effective improvements in health come from toxicologist-physician consultation for complex patients and that telehealth would be more appropriately used to extent generalist access to toxicologists rather than increasing direct patient access to toxicologists.

\addtolength{\textheight}{-.2cm} 

\bibliographystyle{ieeetr}
\bibliography{sample}

\end{document}